# Improving Proton Dose Calculation Accuracy by Using Deep Learning

Chao Wu*[1,2,3], Dan Nguyen[1], Yixun Xing[1], Ana Barragan Montero[3], Jan Schuemann[5], Haijiao Shang[2,3], Yuehu Pu[2], Steve Jiang[1]

[1] Medical Artificial Intelligence and Automation (MAIA) Laboratory, Department of Radiation Oncology, University of Texas Southwestern Medical Center, Dallas, TX, USA
[2] Shanghai Institute of Applied Physics, Chinese Academy of Sciences, Shanghai, China
[3] University of Chinese Academy of Sciences, Beijing, China
[4] Molecular Imaging Radiation Oncology (MIRO) Laboratory, UC Louvain, Brussels, Belgium
[5] Department of Radiation Oncology, Massachusetts General Hospital, Boston, MA, USA

Corresponding Author: Steve.Jiang@UTSouthwestern.Edu

## Abstract

Accurate dose calculation is vitally important for proton therapy. Pencil beam (PB) model-based dose calculation is fast but inaccurate due to the approximation when dealing with inhomogeneities. Monte Carlo (MC) dose calculation is the most accurate method, but it is time consuming. We hypothesize that deep learning methods can boost the accuracy of PB dose calculation to the level of MC. In this work, we developed a deep learning model that converts PB to MC doses for different tumor sites. The proposed model is based on our newly developed hierarchically densely connected U-Net (HD U-Net) network, and it uses the PB dose and patient CT image as inputs to generate the MC dose. We used 290 patients (90 with head and neck, 93 with liver, 75 with prostate, and 32 with lung cancer) to train, validate, and test the model. For each tumor site, we performed four numerical experiments to explore various combinations of training datasets. Training the model on data from all tumor sites together and using the dose distribution of each individual beam as input yielded the best performance for all four tumor sites. The average gamma index (1mm/1% criteria) between the converted dose and the MC dose was 92.8%, 92.7%, 89.7% and 99.6% for head and neck, liver, lung, and prostate test patients, respectively. The average time for dose conversion for a single field was less than 4 seconds. In conclusion, our deep learning-based approach can quickly boost the accuracy of proton PB dose distributions to that of MC dose distributions. The trained model can be readily adapted to new datasets for different tumor sites and from different hospitals through transfer learning. This model can be added as a plug-in to the clinical workflow of proton therapy treatment planning to improve the accuracy of proton dose calculation.

## 1. Introduction

Proton therapy has attracted increasing attention in recent years. The main rationale for using proton therapy is the physical characteristics of the depth dose curve, which has a dose peak (Bragg peak) at a well-defined depth in tissue [1]. This physical advantage allows proton therapy to achieve higher dose conformity in the tumor volume with a lower dose to the surrounding healthy tissue than conventional radiation therapy. Nonetheless, uncertainties in dose calculation have a bigger impact on the desired dose distributions in proton therapy, so accurate dose calculation is essential for the success of a proton therapy treatment [2]. Currently, many pencil beam (PB)-based dose calculation algorithms, based on the works of Hong et al. [3] and Schaffner et al. [4], are widely used in clinical practice. These algorithms provide fast computation but come at the expense of lower dose calculation accuracy in the presence of tissue heterogeneity [5]. This is mainly because these algorithms adopt approximations that disregard lateral inhomogeneities to achieve fast dose calculations [6,7]. In PB dose calculation, the proton dose is computed by using the water equivalent path length along the central path of a pencil beam, and the medium on the central axis is assumed to be laterally infinite and homogeneous [6]. These approximations lead to inaccurate modeling of the multiple Coulomb scattering (MCS) process and, therefore, cause both dose distortion and range uncertainties, especially in the presence of complex geometries and heterogeneous environments [8]. Moreover, PB algorithms' approximations in modeling elastic and inelastic nuclear interactions can also lead to considerable errors in dose calculation, even in homogeneous geometries [6,7].

Monte Carlo (MC) methods are recognized as the gold standard for dose calculation because they can simulate particle propagation through materials by randomly sampling the cross-section of interactions [7-10]. Several phantom studies comparing PB dose calculations and MC algorithms have demonstrated that MC algorithms can provide more accurate dose distributions than PB algorithms [11-14]. For example, in a multi-institution phantom study, Taylor et al reported that the dose distribution obtained with the MC algorithm matched the phantom dose measurement more closely than the dose distribution obtained with the PB algorithm [11]. Besides phantom studies, PB dose accuracy deficiencies have also been investigated in several studies comparing PB dose with MC dose in different tumor sites [8,9,15, 16]. Schuemann et al demonstrated that current PB dose calculation algorithms can cause underdosage to the target by as much as 5%, which can result in differences in tumor control probability of up to 11%. For complex geometries (head and neck cancer and lung cancer) and very deep-seated targets (prostate cancer), MC simulations should be considered instead of PB dose calculations [8]. Moreover, proton therapy can be very sensitive to patient anatomical changes that may distort the planned dose distribution and deteriorate the treatment quality. A recent study has also shown that online adaptation with MC dose calculation can further improve treatment quality for inter-fractional patient geometry changes [17]. Despite all these advantages, MC dose calculation methods have not been widely used in clinical routine because of their computational burden and implementation complexity [7].

Over the years, the research community has devoted significant efforts to accelerating MC dose calculation for proton therapy [18]. Recently, parallel computing techniques based on graphics processing units (GPU) have been employed for this application. Different groups have achieved considerable acceleration factors over conventional CPU-based computations [18-25]. To date, the computation time of the fastest reported MC dose engine is in the minute range. Furthermore, some commercial treatment planning systems (e.g., RayStation and Eclipse) have already included the MC dose calculation feature for proton therapy.

Recently, deep convolutional neural networks have been successfully used to predict patient-specific dose distributions from anatomical information [26-33]. Although these works tried to learn the relationship between patient anatomy and the optimal dose distribution, they did inspire us to use deep learning methods to learn the relationship between tissue inhomogeneities and the differences between PB and MC dose distributions.

In this paper, we present a novel approach that uses deep learning techniques to achieve MC dose calculation accuracy with PB dose calculation efficiency for proton therapy. Specifically, we developed a deep learning model that can precisely and efficiently convert a 3D PB dose distribution to a dose distribution with MC-equivalent accuracy for different tumor sites. The model architecture and training details are presented in Section 2. In Section 3, we present the results of the proposed model. We discuss some future work in Section 4 and draw conclusions in Section 5.

## 2. Methods

### 2.1 Model architecture

The model developed in this work is based on the 3D HD U-Net that was developed and tested for voxel-wise 3D dose prediction for patients with head and neck cancer [28]. HD U-Net combines the essence of two influential neural networks: U-Net [34] and DenseNet [35]. In general, HD U-net's architecture combines U-net's ability to abstract both local and global features from input images with DenseNet's efficient feature propagation and reuse, while maintaining a reasonable memory usage [28,32]. Figure 1 shows the architecture of the HD U-net model modified for this work. The details of the three operations between layers (Dense Convolve, Dense Downsample and U-net Upsample) have been previously introduced elsewhere [28].

The proposed model contains two input channels: one for the 3D proton PB dose distribution and the other for the corresponding CT image. The model has 5 max pooling and 5 upsampling operations, which decrease the image patch size from 128 x 128 x 16 voxels to 8 x 8 x 1 voxels, then increase it back to 128x 128 x 16 voxels. The convolutional kernel size is set to 3 x 3 x 3, and the max pooling size is set to 2 x 2 x 1. Batch normalization is added after the convolution and before the rectified linear unit operations. The dropout rate is set to 0 because no overfitting issue has been observed during training.

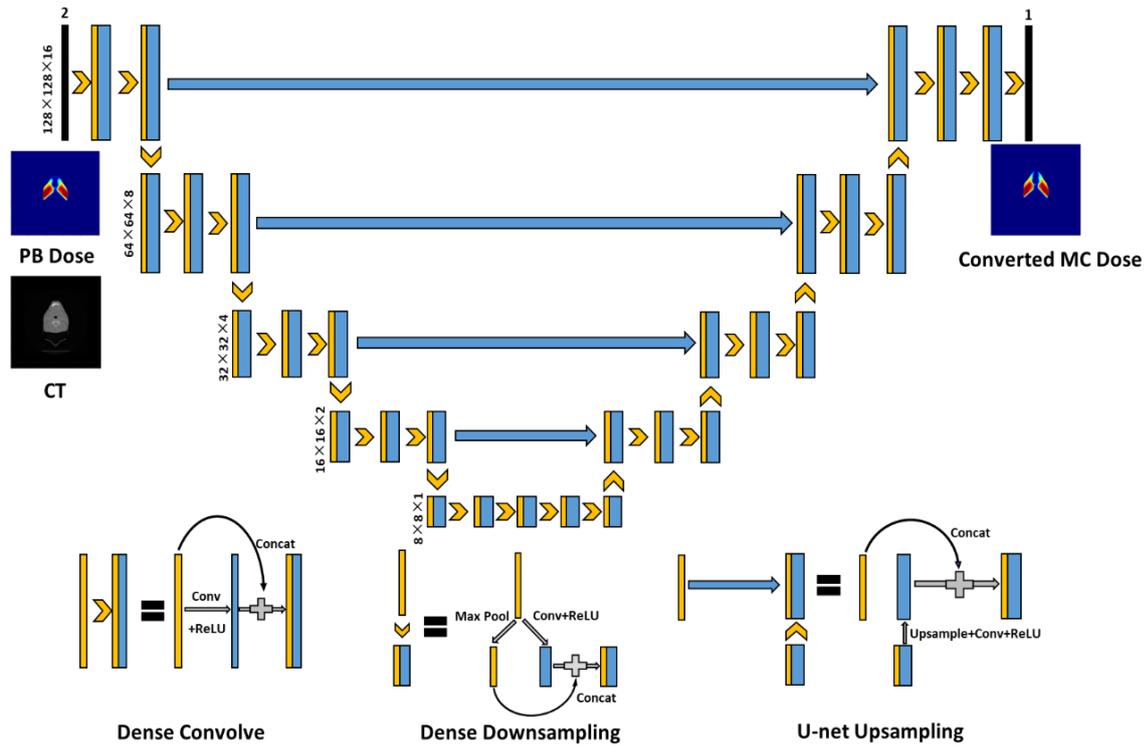

**Figure 1.** Architecture of the HD U-Net. Black numbers on the left side of the model represent the volume shape and resolution at a specific hierarchy. Blue features represent the newly calculated features and trainable parameters to learn. Yellow features are copied or max pooled features that do not need trainable parameters.

### 2.2 Patient database

The patient database used in this work consists of 90 head and neck, 93 liver, 75 prostate and 32 lung patients treated the double scattering method at the Massachusetts General Hospital (MGH) proton therapy center. For each tumor site, about 20% of the patients were randomly selected as the testing set, and the rest were used for training and validation. The specific numbers of test patients and training & validation patients for each tumor site are presented in Table 1. It should be noted that, some treatment plans were only used for research use or as alternate plan options or beam arrangements and the corresponding treatment fields were not actually delivered to patients. For each patient, the proton PB dose was calculated using an algorithm implemented on the XiO treatment planning system (by Computerized Medical Systems Inc, now by ELEKTA), and the MC dose was obtained by using TOPAS version 3.0.1[36], which is based on Geant4.10.3[37]. All the dose data and their corresponding CT images were resampled to have a voxel resolution of 2 x 2 x 2.5 mm$^3$.

**Table 1.** Beam numbers and distribution details of the testing set and the training & validation set for each tumor site.

|  | **Head and Neck** | **Liver** | **Lung** | **Prostate** | **Total** |
|---|---|---|---|---|---|
| Number of patients | 90 | 93 | 32 | 75 | 290 |
| Training & Validation | 72 | 75 | 26 | 61 | 234 |
| Testing | 18 | 18 | 6 | 14 | 56 |
| Number of beams | 720 | 215 | 88 | 260 | 1283 |

*2.3 Training Experiment Design*

The patient data of the four tumor sites used in this work have different beam configurations, so the model must be able to maintain high performance across different beam settings. To address this issue, for each patient, we rotated the dose distribution of each beam and the CT image to the same default angle and then used them as inputs for the model. As the model output, the converted dose distribution of each beam is rotated back and added to the dose distributions of other beams to obtain the total converted dose. In this work, the default angle is set to a 270° gantry angle and a 0° couch angle. We believe this method allows the model to better learn the upstream and downstream characteristics when the proton beam passes through the patient's body and deposits energy along the path, and therefore helps the model to better learn the accurate mapping between the PB dose and the MC dose in relation to tissue inhomogeneities. We also implemented and tested an alternative method that directly uses the composite dose distribution as the model input and output. In this article, we refer to the method that uses the beam dose as the model input and output as Method 1 and the method that uses the composite dose as Method 2. The schematic flow chart of the two methods is shown in Figure 2.

Because the prescription dose varies from patient to patient and from tumor site to tumor site, all the dose and CT images are normalized before being input into the model. For both methods, the CT voxel values are normalized to be between 0 and 1, and the dose distribution voxel values are normalized by dividing the 95th percentile dose value of all voxels receiving dose values greater than 5% of the maximum PB dose. On the output side of the model, the converted beam or composite dose is multiplied by the same normalization factor.

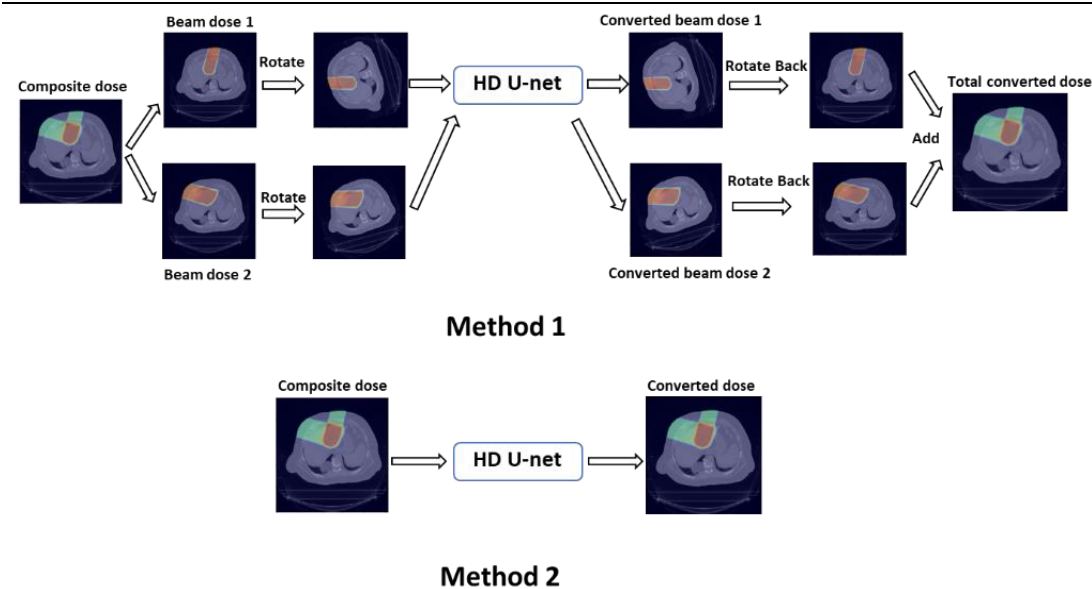

**Method 1**

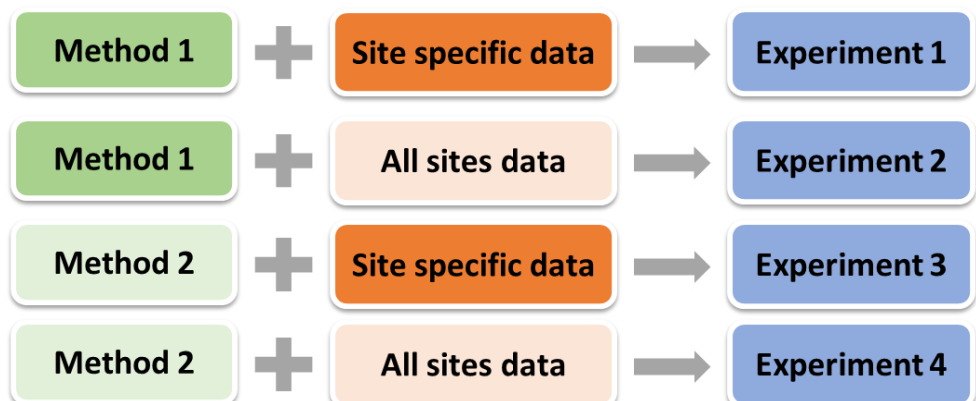

**Method 2**

**Figure 2.** Flow chart of the two training methods adopted in this work.

To develop a general model that can convert PB dose distribution to MC dose distribution for each tumor site, in addition to using site-specific data for training, we implemented joint training that used data from all sites for each method. For each tumor site, four different numerical experiments were carried out to investigate which model had the best performance. Take head and neck patients as an example: for Experiment 1, the beam dose distributions of head and neck patients were used as the model input, and for Experiment 2, the beam dose distributions of all four tumor sites were used as the model input. For Experiment 3, the composite dose distributions of head and neck patients were used as the model input, and for Experiment 4, the composite dose distributions of all four tumor sites were used as the model input. The details of the four experiments are summarized in Figure 3. In total, 10 models were trained and evaluated (two general models trained with all-sites data and eight models trained with site-specific data). For each tumor site, the models of all four experiments were tested on the same test dataset.

| Method 1 | + | Site specific data | → | Experiment 1 |
| Method 1 | + | All sites data | → | Experiment 2 |
| Method 2 | + | Site specific data | → | Experiment 3 |
| Method 2 | + | All sites data | → | Experiment 4 |

**Figure 3.** Schematic overview of the four experiments for each tumor site.

*2.4 Model Training details*

The mean squared error (MSE) between the converted dose distribution and the MC dose distribution was used as the loss function for training each model. The learning rate of each model was adjusted to minimize the validation loss as a function of epochs. During each training iteration, a patch of size 128×128×16 was randomly selected from the patient volume. This patch-wise training method, similar to data augmentation, can reduce overfitting [28]. The Adam algorithm was selected as the optimizer to minimize the loss function. All the deep learning models were built and implemented in Keras with Tensorflow [38] as the back end. Each model was trained for 200 epochs on one NVIDIA Tesla V100 GPU card with 32 GB RAM.

## 3. Results

### 3.1 Gamma index and MSE results

To assess the accuracy of the converted dose, we computed the MSE between the converted dose and the MC dose above a threshold of 10% of the maximum MC dose in the test dataset. We also computed the 3D gamma index ($\gamma$), which can evaluate the dosimetric accuracy of voxels by combining the distance difference and dose difference metrics [39]. The gamma index with 1%/1mm criteria and the MSE results of all experiments for head and neck and liver test patients are presented in Table 2; the corresponding results for lung and prostate test patients are presented in Table 3. Both tables include the PB dose metrics for comparison. For every tumor site, the gamma index and MSE results are substantially better for all the experiments than for the same metrics of the PB dose. Method 1 (Experiments 1 and 2), which uses the beam dose as the model input, clearly outperformed Method 2 (Experiments 3 and 4), which directly uses the composite dose as the model input. Experiment 2, which used the beam dose distribution data from all sites as the model input, clearly outperformed Experiment 1, which used the beam dose distributions of site-specific data as the model input, except in the case of prostate test patients. In general, Experiment 2, which used the beam dose distribution data from all sites as the model input, had the best performance across all four types of test tumor sites. The average gamma index between the converted and the MC dose distributions for head and neck, liver, lung and prostate test patients was 92.8%, 92.7%, 89.7% and 99.6%, respectively.

**Table 2.** Gamma index (1%/1 mm) and MSE results of all the experiments for head and neck and liver test patients (above a threshold of 10% of the maximum MC dose).

|  | Head and Neck | | Liver | |
|---|---|---|---|---|
|  | Gamma Index | MSE(Gy) | Gamma Index | MSE(Gy) |
| **PB dose** | (73.3±6.3) % | (4.89±3.39) | (79.2±5.1) % | (1.72±0.65) |
| **Experiment 1** | (90.4±2.8) % | (1.28±0.92) | (91.9±4.8) % | (0.43±0.31) |
| **Experiment 2** | (92.8±2.9) % | (1.14±0.82) | (92.7±2.9) % | (0.31±0.15) |
| **Experiment 3** | (77.9±9.6) % | (2.64±1.84) | (89.5±4.9) % | (0.66±0.43) |
| **Experiment 4** | (83.0±5.1) % | (2.26±1.58) | (88.9±5.1) % | (0.69±0.42) |

**Table 3.** Gamma index (1%/1 mm) and MSE results of all the experiments for lung and prostate test patients (above a threshold of 10% of the maximum MC dose).

|  | Lung | | Prostate | |
|---|---|---|---|---|
|  | Gamma Index | MSE(Gy) | Gamma Index | MSE(Gy) |
| **PB dose** | (65.4±5.3) % | (3.17±1.80) | (73.3±2.7) % | (2.04±1.10) |
| **Experiment 1** | (86.3±5.9) % | (0.66±0.62) | (99.5±0.3) % | (0.13±0.08) |
| **Experiment 2** | (89.7±3.8) % | (0.48±0.39) | (99.6±0.3) % | (0.12±0.08) |
| **Experiment 3** | (70.3±5.3) % | (2.12±1.66) | (98.6±0.9) % | (0.23±0.16) |
| **Experiment 4** | (75.5±5.2) % | (1.60±0.97) | (97.2±1.4) % | (0.30±0.16) |

*3.2 Dosimetry analysis*

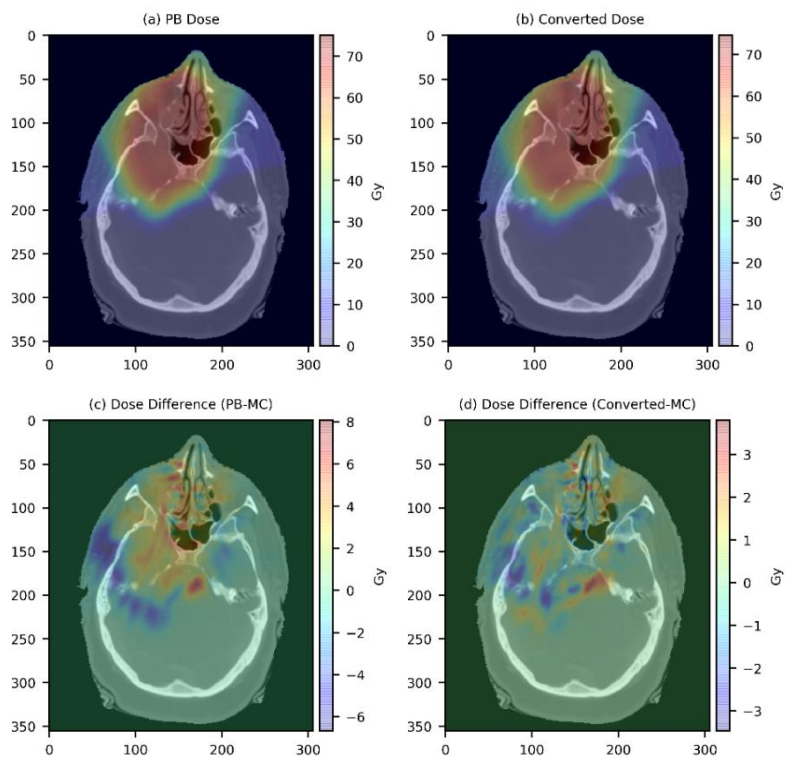

**Figure 4.** Dose color wash of an axial slice close to the center of the target volume for one example head and neck test patient. (a) PB dose distribution; (b) Converted dose distribution; (c) Absolute dose difference between the PB dose distribution and the MC dose distribution (PB-MC), and (d) Absolute dose difference between the Converted dose distribution and the MC dose distribution (Converted-MC).

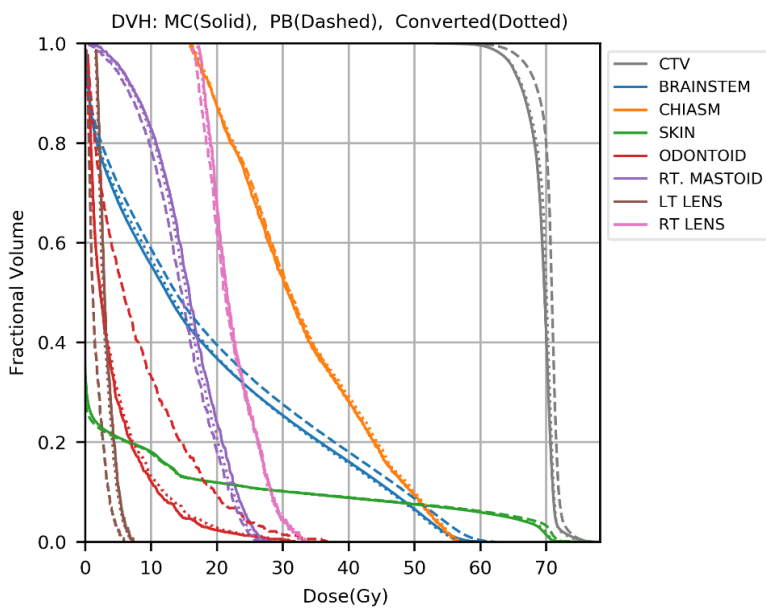

**Figure 5.** DVH plots of the MC dose distribution(solid), the PB dose distribution (dashed), and the converted dose distribution (dotted) for one example head and neck test patient.

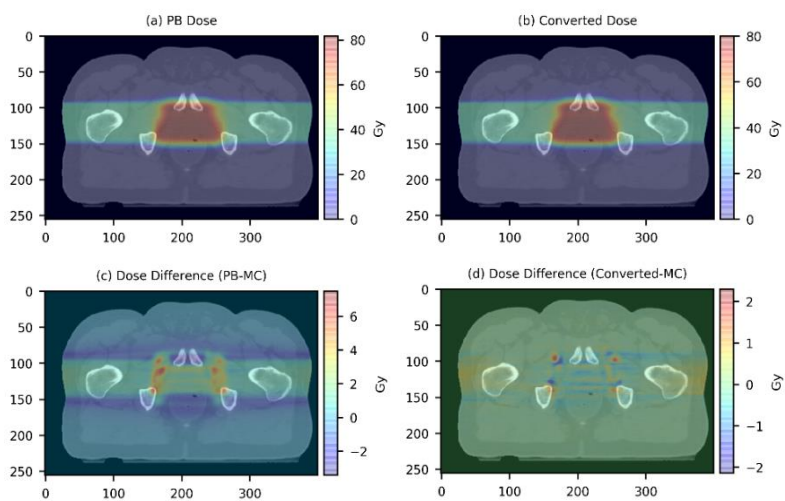

**Figure 6.** Dose color wash of an axial slice close to the center of the target volume for one example prostate test patient. (a) PB dose distribution; (b)Converted dose distribution; (c) Absolute dose difference between the PB dose distribution and the MC dose distribution (PB-MC), and (d) Absolute dose difference between the Converted dose distribution and the MC dose distribution (Converted-MC).

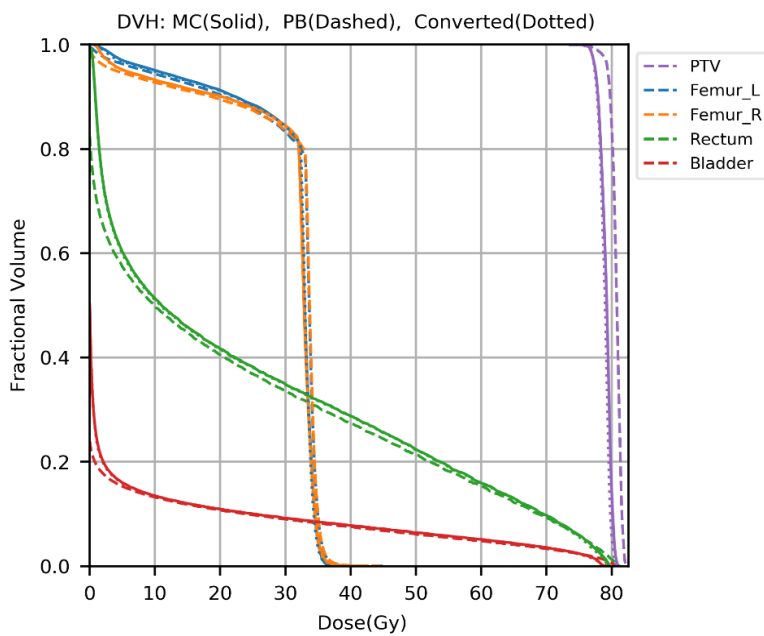

**Figure 7.** DVH plots of the MC dose distribution(solid), the PB dose distribution (dashed), and the converted dose distribution (dotted) for one example prostate test patient.

To illustrate the performance of the developed model, we compared the converted and the MC dose distributions for one example head and neck and prostate test patient. Figures 4 and 6 show dose color washes for one example patient from the head and neck and prostate test sets, respectively. In each of these two figures, one axial slice of the PB dose distribution, the converted dose distribution, the absolute differences between the PB and MC dose distribution and the absolute differences between the converted and MC dose distribution are illustrated in sub-figures a, b, c and d, respectively. Figures 5 and 7 show the dose volume histograms (DVHs)curves of the same example head and neck and prostate test patient, respectively. Upon visual inspection of the dose color washes, the converted dose distributions are almost identical to the MC dose distributions for each example patient. The DVH curves of the converted dose are found to be very close to those of the MC dose for both targets and OARs, but an obvious gap can be observed between the DVH curves of PB and MC doses.

The results of the other test patients and other tumor sites are not presented here because of space limits in the manuscript, but the behaviors are similar for all of them.

We compared the voxel-based dose difference between the PB dose and the MC dose and between the converted dose and the MC dose for all test patients. If any voxel in the PB or the converted dose distribution received more than 3% of the maximum MC dose and its absolute dose difference with the MC dose distribution was larger than 3% of the maximum MC dose, then that voxel was used for evaluation. Dose difference histograms for 18 head and neck, 18 liver, 6 lung and 14 prostate test patients are shown in Figure 8. We can clearly see that, for each type of test patient, the number of voxels with dose differences larger than 3% of the maximum MC dose is substantially lower for the converted dose than for the PB dose.

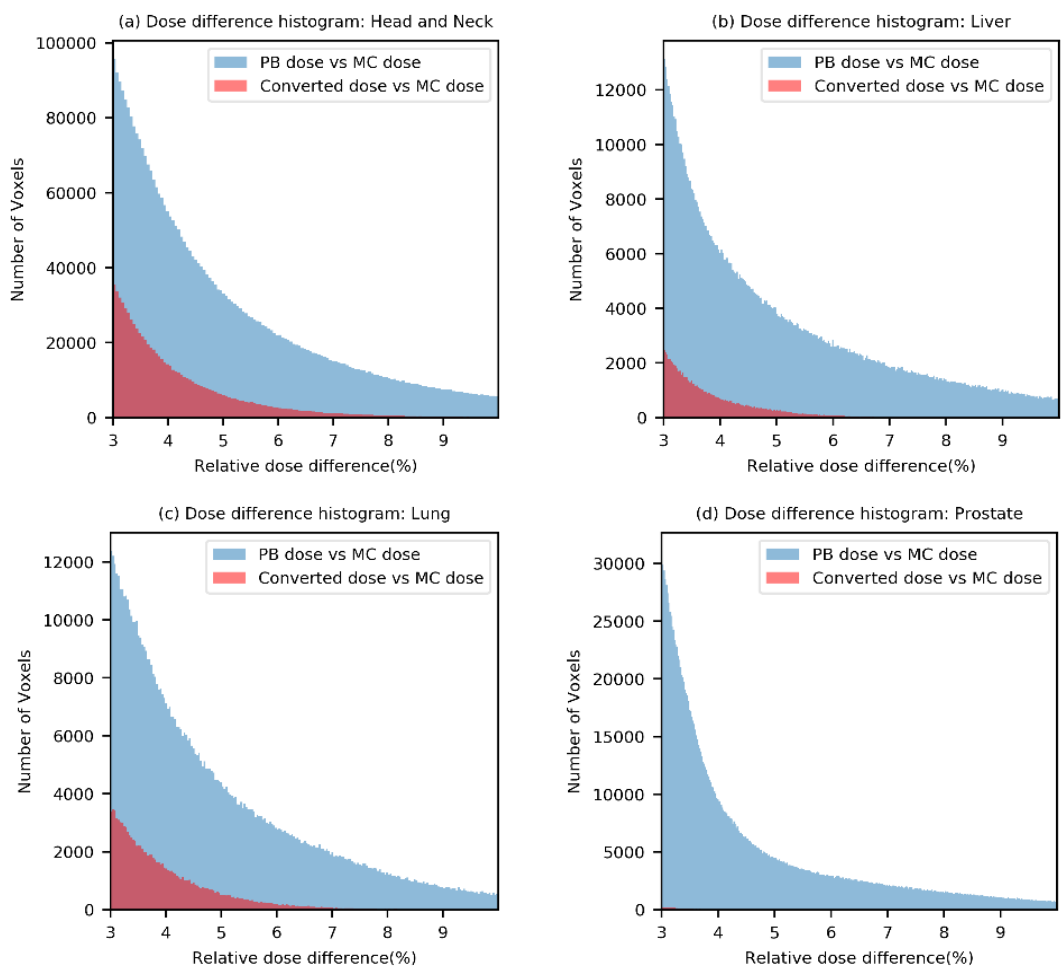

**Figure 8.** Dose difference histograms for all test patients. (a) Head and neck test patients; (b) liver test patients; (c) lung test patients, and (d) prostate test patients.

### 3.3 Efficiency analysis

Table 4 shows the average time required for MC simulation, PB dose calculation and dose conversion of a single field for each tumor site. The MC simulations were implemented in TOPAS, and the PB doses were calculated using an algorithm implemented on the XiO system, both calculations were performed at the MGH proton therapy center. The conversion processes were executed on a NVIDIA Tesla V100 card with 16 GB dedicated RAM. The total time to obtain the converted MC dose is roughly the sum of the PB dose calculation time and the conversion time. Compared with the MC simulation time, the total time to obtain the converted MC dose is substantially shorter.

**Table 4.** Average single-field MC simulation, PB dose calculation, and dose conversion time results for all four tumor sites used in this work (all in seconds).

|  | **Head and Neck** | **Liver** | **Lung** | **Prostate** |
|---|---|---|---|---|
| MC simulation | 14487±12819 | 28652±20041 | 27372±28206 | 77530±25882 |
| PB dose calculation | 123±91 | 253±199 | 303±390 | 172±59 |
| Dose conversion | 3.3±0.8 | 2.2±0.7 | 2.1±0.2 | 2.2±1.0 |

### 3.4 Generalizability analysis

Experiment 2, which used the beam dose distribution data from all sites as the model input, generally gave the best performance. To test the generalizability of this model, we designed and implemented Experiment 5. In this experiment, for a tumor site A, first, we implemented joint training using the beam dose distributions of the other three tumor sites B, C and D as the model input. The model was trained for 200 epochs, and the learning rate was adjusted to minimize the validation loss as a function of epochs. Then, the trained model was fine-tuned by using the beam dose of tumor site A for 100 epochs. The learning rate for fine-tuning was set to 1E-4. After joint training and fine-tuning, the model was tested on the same test dataset of tumor site A as the other experiments. The gamma index and MSE results of Experiments 2 and 5 for the four types of test patients are presented in Table 5, both using MC results as reference. In general, Experiments 2 and 5 had similar performance. The models trained in Experiment 5 showed good performance for each tumor site, with the average gamma index between the converted dose and the MC dose above 90%. These results show that the models that use the beam dose distributions of three tumor sites as input can be easily adapted to a new tumor site, which suggests that joint training using beam dose distributions of multiple tumor sites as the model input has good generalizability. This supports our assumption that the model that uses the beam dose distributions of all four tumor sites as input in Experiment 2 can be adapted to new datasets in clinical practice.

**Table 5.** Gamma index (1%/1 mm) and MSE results of Experiments 2 and 5 for all test patients (above a threshold of 10% of the maximum MC dose).

|  | **Experiment 2** | | **Experiment 5** | |
|---|---|---|---|---|
|  | Gamma Index | MSE (Gy) | Gamma Index | MSE (Gy) |
| **Head and Neck** | (92.8±2.9) % | (1.14±0.82) | (91.9±3.1) % | (1.21±0.82) |
| **Liver** | (92.7±2.9) % | (0.31±0.15) | (93.7±2.8) % | (0.31±0.17) |
| **Lung** | (89.7±3.8) % | (0.48±0.39) | (90.3±4.0) % | (0.46±0.38) |
| **Prostate** | (99.6±0.3) % | (0.12±0.08) | (99.6±0.3) % | (0.11±0.07) |

To further test the generalizability of the model trained in Experiment 2 and to see whether it can be adapted to a new dataset from another hospital, we designed and implemented Experiment 6. The new dataset we used in Experiment 6 includes 27 patients with non-small cell lung cancer from a hospital other than MGH. We randomly selected 22 of these patients as the training set to fine-tune the model and used the remaining 5 patients as the testing set. For each patient, we used RayStation 9A to develop intensity modulated proton therapy (IMPT) treatment plans, and to calculate the PB and MC dose distributions. Method 1, which uses beam dose as the model input, was applied, and the beam dose distributions of the patients in the training set were used to fine-tune the model trained in Experiment 2. We set the total number of epochs to 100 and the learning rate to 1E-4. After fine-tuning, we evaluated the model on the testing set. For the patients in the testing set, the mean gamma index (1%/1 mm) between PB and MC dose distributions was 75.5% ± 8.9% (standard deviation). We found that the converted dose distributions showed substantial improvement over the PB dose distributions and had strong agreement with the MC dose distributions. The mean gamma index between the converted dose and MC dose (1%/1 mm) was 91.8% ± 4.8% (standard deviation). This result proves that the model trained in Experiment 2 can be adapted to a new dataset from different hospitals through simple fine-tuning.

## 4. Discussion

In this work, we present a novel method that uses deep learning methods to convert PB dose distributions to MC dose distributions for different tumor sites. We performed four different experiments to compare different combinations of training datasets, and we found that joint training using the beam dose distributions data from all sites as the model input had the best performance. The converted dose distributions showed substantial improvement over the PB dose distributions in all evaluation criteria that we considered: gamma index, MSE, DVH and dose difference histogram. The developed model can learn from a heterogeneous database that includes beam dose distributions from four tumor sites with different beam configurations and maintain high performance for each tumor site. Using beam dose distributions as the model input (Experiments 1 and 2) yielded substantially better model performance than directly using composite dose distributions as the model input (Experiments 3 and 4). These results suggest that rotating the normalized beam dose and the corresponding CT to the same angle and using them as the model input helps the model to better learn the difference between the PB and the MC dose distributions in relation to tissue heterogeneity and thus improves the model's performance. We also noticed that the converted dose of prostate test patients had the best performance, and the converted dose of lung test patients had the worst, in terms of gamma index and MSE results. One possible reason is that all the prostate patient data used in this work have similar beam settings. The beam angles for prostate patients are set to be either 90° or 270° gantry angle, which we believe makes the prostate patients' data size needed for the model to achieve high performance smaller than other tumor sites. The gamma index and MSE results of Experiments 1 and 2 (see Table 3) show that, for prostate patients, adding the beam dose of other tumor sites for training did not improve the model's performance. This indicates that, in this work, the prostate patient data size is big enough for the model to learn the mapping between the PB and the MC dose distributions. For the other three treatment sites — especially for lung patients which have the smallest data size, adding the training data of the other sites improved the model's performance. Thus, joint training using beam dose distributions from all the tumor sites as the model input can improve the model's performance when the beam configurations are heterogeneous and the data size is relatively small.

The average time needed to convert a single field for all four types of patients in this work is less than 4 seconds. To date, the fastest proton PB dose calculations are in the sub-second range. Da Silva et al achieved PB dose calculations with a double Gaussian kernel in 0.22 s [40]. Because the number of fields used in proton therapy is usually no more than 4, the total time required to obtain the converted dose can potentially be kept within half a minute. This will allow the implementation of online adaption of MC calculations to further improve the treatment quality for inter-fractional geometry changes [17]. We have not yet dedicated any efforts to improving the model's efficiency, which can be easily achieved through methods like model compression; we will explore such methods in our future work. In addition, by accurately and efficiently generating converted MC doses, the proposed model could be used as a plan evaluation or even a decision support tool for physicians.

We have also demonstrated generalizability of the model. In Experiment 5, we showed that the trained model can be easily used for another tumor site through transfer learning. With Experiment 6, we also showed that the trained model can be deployed to different hospitals through transfer learning, even when using a different treatment plan delivery, i.e. IMPT vs. double scattering

The patient database used in this work comprises 90 head and neck, 93 liver, 75 prostate and 32 lung patients treated with double scattering methods at MGH proton center. Currently, IMPT which uses scanned proton pencils to shape the delivered dose distributions, is adopted by most new proton therapy facilities. We plan to extend this work by applying the developed model to IMPT plans of different tumor sites and testing the performance.

## 5. Conclusions

We used a deep learning-based approach to improve the accuracy of proton therapy by converting PB dose distributions to MC-equivalent dose distributions. We carried out four different experiments to explore various combinations of composite versus single beam dose distributions and all-site data versus site-specific data. The results showed that joint training using beam dose distributions of all four tumor sites data available had the best performance. The developed model used in this experiment can accurately and efficiently convert PB dose distributions to MC dose distributions for different tumor sites. The trained model was proved to have good generalizability and can be readily adapted to new datasets for different tumor sites and from different hospitals through transfer learning. This model can be added as a plug-in to the clinical workflow of proton therapy treatment planning to improve the accuracy of proton dose calculation.

**Conflict of interest statement**

None.

**Acknowledgement**

We would like to thank Dr. Jonathan Feinberg for editing the manuscript.